\documentclass[prl,twocolumn,showpacs]{revtex4}
\usepackage{graphicx}
\begin{document}

\title{Orbital ordering in charge transfer insulators}
\author{M. V. Mostovoy and D. I. Khomskii}
\affiliation{Materials Science Center, University of
Groningen, Nijenborgh 4, 9747 AG Groningen, The
Netherlands}

\begin{abstract}
We discuss a new mechanism of orbital ordering, which in
charge transfer insulators is more important than the
usual exchange interactions and which can make the very
type of the ground state of a charge transfer insulator,
i.e.\ its orbital and magnetic ordering, different from
that of a Mott-Hubbard insulator. This purely electronic
mechanism allows us to explain why orbitals in Jahn-Teller
materials typically order at higher temperatures than
spins, and to understand the type of orbital ordering in a
number of materials, e.g.\ K$_2$CuF$_4$, without invoking
the electron-lattice interaction.

\end{abstract}

\date{\today}


\maketitle

Since the seminal Zaanen, Sawatzky, and Allen paper
\cite{ZaanenSawatzkyAllen}, it is widely accepted that
strongly correlated insulators can be divided into two
classes according to the nature of their gap. The
Mott-Hubbard insulators (MHI) are characterized by the
lowest-energy excitations of the type $d^n + d^n
\rightarrow d^{n-1} + d^{n+1}$, obtained by transferring a
$d$-electron from one transition metal (TM) ion to
another, whereas in charge transfer insulators (CTI) the
corresponding excitations are holes on ligand sites
(oxygen ions in the case of TM oxides): $d^n + p^6
\rightarrow d^{n+1} + p^5$. MHI are usually described by
an effective Hubbard model, in which only the $d$-electron
states are considered explicitly. On the other hand, CTI
are described by models that include also the relevant
oxygen $p$-states, such as the $dp$ (or three-band
Hubbard) model of high-$T_c$ superconductors \cite{Emery}.
The value of the gap is determined by the smallest of the
two parameters: the energy of electron transfer from
ligand to TM sites $\Delta$ and the Hubbard repulsion on
TM sites $U_d$. For CTI (usually, the oxides of heavier
$3d$ metals) $\Delta < U_d$, while for MHI $U_d < \Delta$.

The distinction between these two types of insulators
becomes apparent upon doping and may play an important
role in high-$T_c$ superconductors \cite{Emery,Varma},
though according to other point of view,  there is no
qualitative difference between doped CTI and MHI
\cite{ZhangRice}. In the absence of doping, both CTI and
MHI with one electron or hole per TM site are $S =
\frac{1}{2}$ Heisenberg antiferromagnets.

The situation turns out to be quite different in
Jahn-Teller systems, in which $d$-electrons, in addition
to spins, have also orbital degrees of freedom.  In this
Letter we show that in insulators with orbital degeneracy
not only the excitation spectrum, but also an orbital and
magnetic ordering in the ground state may strongly depend
on the ratio $\Delta / U_d$.  The difference between the
ground states of CTI and MHI results from a peculiar
exchange interaction between orbitals on nearest-neighbor
$d$-sites, which may be called `the orbital Casimir
force', as it originates from the vacuum energy of charge
fluctuations due to the hopping of electrons between the
ligand and TM ions (whereas the usual exchange involves
the electron transfer from one TM site to another). We
recently noted the importance of such interaction in
geometrically frustrated systems with the $90^\circ$
metal-oxygen-metal bonds, {\em e.g.} NaNiO$_2$, where it
determines the type of orbital ordering
\cite{MostovoyKhomskii}. Here we show that this orbital
exchange of electronic origin is generally present in all
TM oxides and, in particular, in materials with the
$180^\circ$ metal-oxygen-metal bonds, like perovskites
ABO$_3$ and their layered analogs A$_2$BO$_4$, and it is
not included in the widely used exchange model
\cite{KugelKhomskiiJETP,KugelKhomskiiUFN} (sometimes
called Kugel-Khomskii (KK) model). That model is derived
from an effective Hubbard model describing only the
degenerate $d$-electron orbitals and which, strictly
speaking, only applies to MHI with $\Delta \ll U_d$.  The
new orbital interaction obtained below helps to resolve a
difficulty of the KK model, in which an orbital ordering
is strongly coupled to a magnetic one, implying close
temperatures of orbital and magnetic orderings
\cite{KugelKhomskiiUFN}, whereas in reality orbitals
usually order at higher temperatures than spins.

\noindent{\em Exchange interactions:}  We first calculate
an exchange Hamiltonian for two  TM ions with one hole on
the doubly degenerate $e_g$-level, {\em e.g.} Cu$^{2+}$
($t_{2g}^6e_g^3$).  The two orthogonal $e_g$-orbitals
$|3z^2 - r^2\rangle$ and $|x^2 - y^2\rangle$ may be
described by, respectively, the up and down states of the
isospin-$\frac{1}{2}$ operator $T^z$
\cite{KugelKhomskiiJETP,KugelKhomskiiUFN}.  It is
convenient to introduce an overcomplete set of operators
$d_{i\alpha\sigma}$ on each TM site $i$ annihiliating a
hole in the state $| 3\alpha^2 - r^2\rangle$ ($\alpha =
x,y,z$) with the spin projection $\sigma$. These states
are hybridized with the $p_\alpha$ orbitals on two
neighboring oxygen sites $i \pm \frac{\alpha}{2}$ (the
corresponding oxygen hole operators are denoted by
$p_{i+\frac{\alpha}{2},\sigma}$). The hole states on TM
sites are described by the standard atomic Hamiltonian,
which includes the on-site repulsion $U_d$ and the Hund's
rule exchange coupling $J_d$ for holes on two orthogonal
orbitals. The Hamiltonian of oxygen holes hybridized with
the $d$-states is
\begin{eqnarray}
H_p &=& H_{kin} + U_p\sum_{i,\alpha}
p_{i+\frac{\alpha}{2}\uparrow}^\dagger
p_{i+\frac{\alpha}{2}\uparrow}
p_{i+\frac{\alpha}{2}\downarrow}^\dagger
p_{i+\frac{\alpha}{2}\downarrow} \nonumber \\&-& t_{pd}
\sum_{i,\alpha,\sigma} \left[d_{i\alpha\sigma}^\dagger
\left( p_{i+\frac{\alpha}{2},\sigma} +
p_{i-\frac{\alpha}{2},\sigma}\right) + h.c. \right]
\label{Hp}
\end{eqnarray}
where the first term is the kinetic energy of holes in the
oxygen $p$-band, the second term is the Hubbard repulsion
on oxygen sites, and the last term describes the hopping
between oxygen and TM sites.

\begin{figure}
\centering
\includegraphics[width=6.5cm]{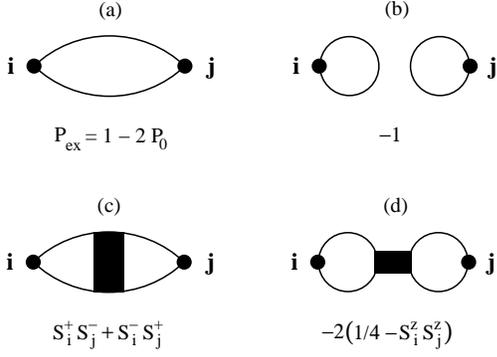}
\caption{\label{diagrams} Diagrams describing exchange
interactions mediated by two virtual oxygen holes.}
\end{figure}

Integrating the oxygen degrees of freedom out, one obtains
an effective interaction between $d$-electrons in the form
of an expansion in powers of $(t_{pd}/\Delta)^2$.  The
lowest order term provides a hopping between TM sites,
which together with the Coulomb and Hund's rule
interactions between $d$-electrons gives an effective
$dd$-model.  For dispersionless oxygen holes, $H_{kin} =
\Delta \sum_{i,\alpha,\sigma}
p^{\dagger}_{i+\frac{\alpha}{2} \sigma}
p_{i+\frac{\alpha}{2} \sigma}$, the effective $dd$-hopping
occurs between nearest-neighbor TM sites, and its
amplitude is $t_{dd} = t_{pd}^2/\Delta$.

The next term in this expansion ($\propto {t_{pd}}^4$)
gives an orbital and spin exchange mediated by two oxygen
holes. The corresponding diagrams are shown in
Fig.\ref{diagrams}, where in the diagrams (a) and (b) two
holes do not interact with each other in the virtual
state, while the black squares on diagrams (c) and (d)
correspond to one or more Hubbard interactions on oxygen
sites.  Each diagram gives rise to a particular form of
spin and orbital exchange.  Thus, if the holes on the TM
sites $i$ and $j$ exchange places [Fig.\ref{diagrams}(a)],
the spin exchange is described by operator $P_{ex} = +1$
if the total spin of the two holes $S = 1$, and $-1$ for
$S = 0$. This operator can also be written in the form
$P_{ex} = 1 - 2 P_0$, where $P_0 = 1/4 - \left({\bf S}_i
{\bf S}_j \right)$ is the projector on the $S = 0$ state.
If, on the other hand, the holes return back to their
starting points, the corresponding `spin factor' is $-1$.
The interacting holes in the diagrams (c) and (d) must
have opposite spin projections, so the spin exchange in
that case is described by $S_i^+ S_j^- + S_i^- S_j^+$ [see
Fig.\ref{diagrams}(c)], while if the spin projections
remain unchanged [see Fig.\ref{diagrams}(d)], the spin
operator is $-2 \left(1/4 - S_i^z S_j^z \right)$.  Since
the diagrams (c) and (d) only differ by the spin factors,
their sum is proportional to $- 2 P_0$, corresponding to
the fact that the on-site Hubbard interaction between two
holes is only nonzero when their spins add to the total
spin $0$.

While the spin-dependence of the exchange interactions,
discussed above, is very general, the orbital structure
depends on the dispersion of holes and on relative
positions of the sites $i$ and $j$. In particular, for
dispersionless holes the exchange in order $t_{pd}^4$ only
occurs between two nearest neighbor TM sites, e.g. $i$ and
$i + \alpha$, and the diagrams (a), (c), and (d) are
proportional to ${\cal I}_{i\alpha} = \left(\frac{1}{2} +
I_i^\alpha\right) \left(\frac{1}{2} +
I_{i+\alpha}^\alpha\right),$ where
\begin{equation}
I^x =  - \frac{1}{2} T^z - \frac{\sqrt{3}}{2} T^x, I^y  =
- \frac{1}{2} T^z + \frac{\sqrt{3}}{2} T^x, I^z = T^z,
\end{equation}
and the operator $\left(\frac{1}{2} + I_i^\alpha\right) =
d_{i\alpha}^{\dagger} d_{i\alpha}$ is the projector on the
state $|3 \alpha^2 - r^2 \rangle$ on the site $i$.  The
contribution of the diagram in Fig.\ref{diagrams}(b) is
proportional to $4 \sum_{\beta,\gamma}\left(\frac{1}{2} +
I_i^\beta\right) \left(\frac{1}{2} + I_{j}^\gamma\right) =
9$, since $I^x+I^y+I^z = 0$.

Combining terms containing $P_0$ with the exchange
interactions appearing in the dd-model in the order
$t_{pd}^4$, we obtain
\begin{eqnarray}
H_{ST} = - \frac{4 t_{pd}^4}{\Delta^2} \sum_{i,\alpha}
\left[ \left(\frac{1}{U_d} + \frac{2}{2 \Delta +
U_p}\right) \right. {\cal I}_{i\alpha} P_0 \nonumber \\+
\frac{J_d}{2 U_d^2} \left. \left(\frac{1}{4} - I_i^\alpha
I_{i+\alpha}^\alpha \right) P_1 \right], \label{HST}
\end{eqnarray}
where the last term describes the Hund's rule coupling on
TM sites and $P_1 = 1 - P_0 = 3/4 + \left({\bf S}_i {\bf
S}_j \right)$ is the projector on the $S=1$ state. In
addition to the coupled spin and orbital interactions,
there exists also a spin-independent orbital exchange
\begin{eqnarray}
H_T^{(h)} &=& \frac{2t_{pd}^4}{\Delta^2}  \sum_{i,\alpha}
\bigg[ \frac{1}{\Delta} \left(\frac{1}{2} +
I_i^{\alpha}\right) \left(\frac{1}{2} +
I_{i+\alpha}^{\alpha}\right) \nonumber
\\ &-&  \frac{1}{U_d} \left(\frac{1}{4} - I_i^\alpha
I_{i+\alpha}^\alpha\right) \bigg] \label{HT}
\end{eqnarray}
The first term is the spin-independent contribution of the
processes shown in Figs.\ref{diagrams}(a) and (b), while
the second one results from the $dd$-model. It is
important to note that in the order $t_{pd}^4$ it is
impossible to exchange two holes in such a way that they
never occupy the same oxygen site (the corresponding terms
in the diagrams (a) and (b) cancel each other). The
orbital exchange interaction containing only powers of
$\Delta$ is actually due to the virtual hopping of holes
to all neighboring oxygen sites and back
[Fig.\ref{diagrams}(b)], except for the oxygen bridging
the two neighboring TM sites. In the limit $\Delta \gg
U_d$ Eqs.\ (\ref{HST}) and (\ref{HT}) reduce to the KK
model.

\begin{figure}
\centering
\includegraphics[width=6.5cm]{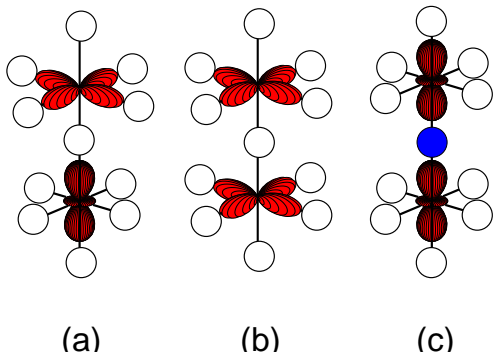}
\caption{\label{pair} Favorable [(a) and (b)] and
unfavorable (c) occupations of hole orbitals of
neighboring TM ions in CTI, in which the exchange is
dominated by Eq.(\ref{HT}) (the terms $\propto U_d^{-1}$
stabilize the state (a)).}
\end{figure}

The term in $H_T$ containing only $\Delta$ in denominator
is similar to the pure orbital exchange for the $90^\circ$
metal-oxygen-metal bonds \cite{MostovoyKhomskii}, and its
origin can be better understood by noting that it is the
only term that does not vanish in the limit of infinite
on-site Hubbard repulsion $U_p$ and $U_d$. It represents
the energy of the delocalization of holes over the
octahedra of the ligand ions, surrounding the TM sites, or
more correctly, the increase of that energy due to the
fact that the neighboring octahedra share one ligand ion
and for infinite $U_p$ the intermediate two-hole states on
the shared ion are forbidden. This energy increase depends
on orbital states of the neighboring holes, which gives
rise to the orbital exchange with the positive sign. Thus
the origin of this orbital interaction is similar to that
of the Casimir effect \cite{Casimir}, resulting from the
dependence of the energy of zero-point fluctuations of
electromagnetic field on positions of two metallic plates.
It is also important that this orbital exchange has
positive sign, while all other terms in Eqs.~(\ref{HST})
and (\ref{HT}) are negative (since they describe the
second-order corrections to the ground state energy in the
effective $dd$-hoping). The result of this new orbital
interaction can be formulated as the following ``rule of
thumb'' (applied also to materials with one electron per
TM site): to avoid the energetically unfavorable
occupation of the bridging oxygen by two holes, the {\em
d-holes} should occupy orbitals directed {\em away} from
the common oxygen either on both neighboring TM sites [see
Fig.\ref{pair}(b)] or, at least, on one site [see
Fig.\ref{pair}(a)].

The importance of the orbital exchange Eq.(\ref{HT})  for
an orbital ordering may be understood already by
considering the minimal-energy states of a pair of holes
on the neighboring TM sites $i$ and $i + z$. In MHI
($\Delta > U_d$) the exchange is dominated by $H_{ST}$
(\ref{HST}) and the last term in Eq.(\ref{HT}). The AFM
exchange between spins of the holes, described by this
term, is the strongest for $I^z = \frac{1}{2}$ on both
sites, i.e., the lowest energy is achieved if both holes
occupy the $|3z^2 - r^2 \rangle$ state, as shown  in
Fig.\ref{pair}(c). The ground state of the pair is a spin
singlet. On the other hand, in CTI ($\Delta < U_d$) the
strongest interaction is the orbital exchange $H_T$ (see
Eq.(\ref{HT})), and the state shown in Fig.\ref{pair}(c)
has the highest orbital energy, since both hole orbitals
are oriented towards the common ligand ion. The more
energetically favorable states in CTI will be those, in
which at least one hole is in the $|x^2 - y^2\rangle$
state (see Fig.\ref{pair}a and b). Furthermore, both the
last term in Eq.(\ref{HT}) and the relatively weak FM spin
exchange due to the Hund's rule coupling select the state
shown in Fig.\ref{pair}a with the total spin of the pair
$S = 1$. Thus, the ground state of the pair in the CT
regime {\em is different} from that in the MH regime.

Consider now two neighboring TM sites with one $e_g$
electron, as {\em e.g.} Mn$^{3+}$ ($t_{2g}^3e_g^1$),
Cr$^{2+}$ ($t_{2g}^3e_g^1$), and the low-spin Ni$^{3+}$
($t_{2g}^6e_g^1$)).  In the KK model the systems with 1
electron/site and 1 hole/site order in exactly the same
way, since the Hubbard model, from which the KK exchange
Hamiltonian is derived, has the electron-hole symmetry
\cite{KugelKhomskiiUFN}; consequently, the term $H_{ST}$
(\ref{HST}) has the same form both for electrons and
holes. On the other hand, in the $dp$-model this symmetry
is lost: the new orbital exchange interaction for
one-electron case has the form [cf. Eq.(\ref{HT})]
\begin{eqnarray}
H_T^{(e)}  &=&   + \frac{2 t_{pd}^4}{\Delta^2}
\sum_{i,\alpha}\bigg[\frac{1}{\Delta}\left(\frac{3}{2} -
I_i^{\alpha}\right) \left(\frac{3}{2} -
I_{i+\alpha}^{\alpha}\right) \nonumber
\\ &-&  \frac{1}{U_d} \left(\frac{1}{4} - I_i^\alpha
I_{i+\alpha}^\alpha\right) \bigg] \label{He1}
\end{eqnarray}
In this case for both MHI and CTI the energy of a pair of
TM electrons on the sites $i$ and $i+ z$ is minimal if
both electrons occupy the $d_{3z^2 - r^2}$ orbital,
Fig.\ref{pair}(c), and their spins form the singlet state.

Note that the electron-phonon (Jahn-Teller) interaction
also gives rise to a purely orbital exchange of the form
$H_{JT} \propto \sum_{i,\alpha} I_i^{\alpha}
I_{i+\alpha}^{\alpha}$, which for both the electron and
hole cases favors the antiferro-orbital (AFO) spin-triplet
state shown in Fig.\ref{pair}(a). Thus the JT interaction
favors the same state as the novel orbital exchange for
one-hole systems, but a different one for one-electron
systems. This, in principle, gives a possibility to
experimentally discriminate between these two mechanisms
and to study their relative importance. The Cu$^{2+}$
pairs (one hole) in K$_2$ZnF$_4$ were found to have the
triplet ground state \cite{Eremin}, showing that the
orbital exchange Eq.(\ref{HT}) and/or JT interaction are
in this case stronger than the KK exchange. We are not
aware of similar data for pairs of one-electron ions.

\noindent{\em Orbital and magnetic ordering:} Equations
(\ref{HST}--\ref{He1}) allow us to consider types of
ordering in different situations.  Thus, e.g., the ground
state of a chain of TM ions in the $z$-direction with 1
hole/site is AFO with alternating $d_{3z^2 - r^2}$ and
$d_{x^2 - y^2}$ orbitals and FM for CTI, as that is
favored by the sum of Eqs.~(\ref{HST}) and (\ref{HT}),
while for MHI, in which $H_{ST}$ dominates, we obtain the
ferro-orbital ordering with $d_{3z^2 - r^2}$ orbitals
occupied on all sites, and the AFM spin
exchange~\cite{lowd}. The latter state also minimizes the
exchange energy of one-dimensional CTI with 1
electron/site.

Similarly e.g.\ for CTI ($U_d \rightarrow\infty$) with a
square lattice and 1 hole/site, the orbital exchange
interaction Eq.(\ref{HT}) has the form
\begin{equation}
H_T  =   + \frac{2 t^4}{\Delta^3}
\sum_{i}\left(I_i^{x}I_{i+x}^{x} + I_i^{y}I_{i+y}^{y} -
I_i^z \right). \label{eq:Hh12d}
\end{equation}
The term bi-quadratic in isospin operators favors an
antiparallel orientation of neighboring isospins, while
the linear term in Eq.(\ref{eq:Hh12d}) acts as `the
orbital field' in the $z$ direction, favoring the canted
orbital state with the canting angle $\pm 60^\circ$, which
corresponds to the two-sublattice orbital ordering with
the $d_{z^2 - x^2}$ and $d_{z^2 - y^2}$ hole states. Since
these two neighboring orbitals do not overlap, the Hund's
rule coupling (the last term in Eq.(\ref{HST})) results in
the FM exchange between the spins of the holes. This is
precisely the orbital and magnetic structure of the
layered compound K$_2$CuCl$_4$.   For MHI, however, we
would get from Eq.(\ref{HST}) a different ground state:
ferro-orbitally ordered with $I^z=-\frac12$ ($d_{x^2 -
y^2}$-orbitals) and AFM spin ordering (see, however,
Ref.~\onlinecite{lowd}). Previously, the KK model was
successfully used to predict the orbital structure of that
compound \cite{KhomskiiKugel}, {\it assuming} that spins
order ferromagnetically; but as we mentioned above, such a
state is not the ground state of the KK model.

The case of a cubic lattice of TM ions is somewhat special
for several reasons. First, using $\sum_\alpha I_i^\alpha
= 0$, we find that the exchange interactions have the
electron-hole symmetry. In particular, both Eq.(\ref{HT})
and Eq.(\ref{He1}) can be written in the form
\begin{equation}
H_T =   + \frac{2 t^4}{\Delta^2}\left(\frac{1}{\Delta} +
\frac{1}{U_d}\right) \sum_{i}\left(I_i^{x}I_{i+x}^{x} +
I_i^{y}I_{i+y}^{y} + I_i^{z}I_{i+z}^{z}\right).
\label{eq:H1}
\end{equation}
Second, the orbital interaction Eq.(\ref{eq:H1}) is
frustrated \cite{FeinerOlesZaanen,GKhaliullin}: apart from
the AFO mean-field ground states with the average isospin
vector $\langle {\bf T}_i \rangle$ parallel (antiparallel)
to an arbitrary vector in the $(T^x,T^z)$-plane, there are
infinitely many disordered ground states, obtained from
the AFO states by performing the transformation $I^\alpha
\rightarrow I^\alpha$, $I^\beta \leftrightarrow I^\gamma$,
where $\alpha$, $\beta$, and $\gamma$ are three different
spatial indexes, on an arbitrary set of planes transverse
to the $\alpha$ axis. This ground state degeneracy is
similar to the one of the orbital Hamiltonian of $e_g$
electrons on a triangular lattice \cite{MostovoyKhomskii}.
The quantum orbital fluctuations select the AFO states
$I^\alpha = \pm \frac{1}{2}$, where $\alpha = x,y$ or $z$,
which have the softest (two-dimensional) orbital
excitations \cite{GKhaliullin}. For instance, for $\alpha
= z$, such a state corresponds to the occupation of
$d_{3z^2 - r^2}$ and $d_{x^2 - y^2}$ orbitals on A and B
sublattices. The corresponding magnetic ordering would be
AFM in the $xy$ planes and FM in the $z$ direction.

If we now would increase the ratio $\Delta / U_d$, moving
from CTI to MHI, the coupling between orbitals and spins
would ultimately stabilize the canted AFO state with an
AFM coupling along the chains in the $z$ direction and a
FM exchange between spins in the $xy$ layers. The detailed
treatment of this case, relevant e.g. for the perovskites
KCuF$_3$ and LaMnO$_3$, will be given elsewhere.

In conclusion, we showed that the description and
properties of orbital ordering in charge transfer
insulators are much different from those in Mott-Hubbard
insulators. In particular, even in perovskite CTI with the
$180^\circ$-exchange the orbital interactions are stronger
than the spin exchange, so that an orbital ordering, in
general, occurs at higher temperatures.  In this respect
the exchange mechanism of an orbital ordering in CTI
resembles the ordering due to the JT (electron-phonon)
interaction, although is not identical to it.  Another
difference between the CTI and MHI is that the
electron-hole symmetry existing in the MH case is lost in
the CT regime, so that the type of an orbital and spin
ordering may be different for one-electron and one-hole
materials. The effective exchange Hamiltonian for CTI
allows us to describe the ordering in a number of such
materials, without invoking any extra factors.  Most
importantly, in orbitally degenerate insulators not only
the excitation spectrum, but also the very type of the
ground state may be different in Mott-Hubbard and charge
transfer regimes, in contrast to the non-degenerate case,
in which it is the same.

This work was supported by the MSC$^{\it plus}$ program
and the Stichting voor Fundamenteel Onderzoek der Materie
(FOM).


\begin{thebibliography}{99}
\bibitem{lowd}These conlcusions may be modified, if we include
anharmonicity effects, considered e.g.\ by D.~I.~Khomskii
and J.~van den Brink, Phys. Rev. Lett. {\bf 85}, 3329
(2000).  In low-dimensional systems the orbital degeneracy
may be also lifted by the symmetry of the crystal
structure, which results in the term $H^\prime = \lambda
\sum_{i} I_i^z$, for the one-dimensional chains along the
$z$ axis and for the two-dimensional $xy$ planes.
\end{thebibliography}
\end{document}